\documentclass[12pt]{elsarticle}

\usepackage{graphics}  
\usepackage{epsfig}
 \usepackage{verbatim}
 \usepackage{subfigure}

\IfFileExists{srcltx.sty}{\usepackage[active]{srcltx}}

\newcommand{\be}{\begin{equation}}
\newcommand{\ee}{\end{equation}}
\newcommand{\bea}{\begin{eqnarray}}
\newcommand{\eea}{\end{eqnarray}}

\begin{document}

\title{Determination of intergalactic magnetic fields from gamma ray data}

\author[ucla]{Warren Essey}
\address[ucla]{Department of Physics and Astronomy, University of California, Los Angeles, CA 90095-1547, USA}
\author[caltech]{Shin'ichiro Ando}
\address[caltech]{California Institute of Technology, Mail Code 350-17, Pasadena, CA 91125, USA}
\author[ucla,ipmu]{Alexander Kusenko}
\address[ipmu]{IPMU, University of Tokyo, Kashiwa, Chiba 277-8568, Japan}



\begin{abstract}
We report a measurement of intergalactic magnetic fields using combined data from Atmospheric Cherenkov Telescopes and Fermi Gamma-Ray Space Telescope, based on the spectral data alone.  If blazars are assumed to produce both gamma rays and cosmic rays, the observed spectra are not sensitive to the intrinsic spectrum of the  source, because, for a distant blazar, secondary photons produced along the line of sight dominate the signal.  In this case, we find $1\times 10^{-17}~{\rm G} < B <  3\times 10^{-14}~{\rm G}$.  If one excludes the cosmic-ray component, the $10^{-17}$~G lower limit remains, but the upper limit depends on the spectral properties of the source.  We present the allowed ranges for a variety of model parameters.  
\end{abstract}

\maketitle

Intergalactic magnetic fields (IGMFs) play an important role in many astrophysical processes and may offer a new window on the cosmology in the early universe, but the size and origin of these fields are still poorly understood~\cite{Kronberg:1993vk}. Until recently, only the upper limits of $10^{-9}$~G were inferred from the observational data~\cite{Barrow:1997mj}.  
One can measure IGMFs using gamma-ray observations, for example, using time delays~\cite{1995Natur.374..430P}, or by searching extended halos around the point objects~\cite{Aharonian:1993vz,Ando:2010rb}.
Here we employ an independent and complementary approach to obtain new upper and lower limits on IGMFs based on the spectra of three blazars observed by HESS, for which there also exist upper limits from Fermi Gamma-Ray Space Telescope. We have calculated the spectra numerically, and we use the goodness of fit to the data as the means to infer the average  values of IGMFs.  Blazar spectra have been used to set {\em lower} bounds on the IGMF~\cite{Neronov:1900zz,Tavecchio:2010mk,Dolag:2010ni,Dermer:2010mm}. However, in each of Refs.~\cite{Neronov:1900zz,Tavecchio:2010mk,Dolag:2010ni}, only one EBL model and a single power law injection spectrum were considered, and the cosmic ray contribution was not included in Refs.~\cite{Neronov:1900zz,Tavecchio:2010mk,Dolag:2010ni,Dermer:2010mm}, while it can be the dominant contribution for distant sources~\cite{Essey:2009zg,Essey:2009ju,Essey:2010er}. Our analysis is different from previous work in that we have scanned over a very wide range of model parameters, and we included the cosmic-ray component. 

AGN are known to be sources of very high energy (VHE) gamma rays; they
are also believed to produce cosmic rays~\cite{Biermann:2009cz,stanev}. There is growing evidence that
both primary gamma rays emitted at the source and secondary gamma rays
produced by cosmic rays along the line of sight contribute to the
observed signals~\cite{Essey:2009zg,Essey:2009ju,Essey:2010er}.  We,
therefore, consider two cases: (i) pure gamma-ray emission at the
source, and (ii) mixed gamma-ray and cosmic ray emission.  In the case
(i) the intrinsic source spectra determine the observed spectra.  If,
however, the cosmic-ray contribution (ii) is included, the observed
gamma-ray spectra are remarkably independent of the intrinsic gamma-ray
or cosmic-ray spectra at the source.   The observed spectrum is
determined almost entirely by the shape (but not the overall
normalization) of extragalactic background light (EBL) spectrum in case
(ii), and the overall agreement of such a prediction with observed
spectra of distant blazars can be considered evidence of cosmic ray
contribution, and, therefore, of cosmic ray acceleration in
AGN~\cite{Essey:2009zg,Essey:2009ju,Essey:2010er}.  Nevertheless, in
what follows, we will consider both possibilities (i) and (ii). In the case (i), most of the observed secondary gamma rays are produced within tens of megaparsecs from the source, and 
our measurement reflects IGMFs closer to the source.   In the case (ii), the field strength we infer corresponds to equally weighted averaging along the line of sight. 

To probe IGMF we constructed a detailed Monte Carlo simulation that accurately tracks the effects of both electromagnetic showers and cosmic-ray propagation and interactions. Included in the simulation are both pion photo-production and proton pair production for cosmic rays and pair production and inverse Compton scattering for gamma rays and electrons. Any relevant secondary decays such as neutron and pion decays were also included. Deflections due to the IGMF are tracked with both momentum and position vectors stored at all points of the propagation. To construct observed spectra, the position and momentum vectors of gamma rays at the $z=0$ surface are used to calculate the image size. Individual gamma-rays are tracked to the instrument, and their arrival
direction further scattered by the instrumental PSF. Any events with arrival directions remaining with the PSF radius are then retained.  (A study of the source morphology as a function of energy 
can be a good diagnostic technique, but in this paper we concentrate only on the spectral analysis.)  For $E_\gamma<100$~GeV the publicly available Fermi PSF was used for comparison and for $E_\gamma>100$~GeV HESS PSF was used. 
 
Suppression of the low-energy spectra of point images depends on the mean field along the line of sight more than on the (poorly known) distribution of IGMFs. 
Indeed, one can use the results of Ref.~\cite{Dolag:2010ni} to compare the effects of uniform magnetic fields with those of more realistic simulations, which take into account 
the distribution of matter in the universe~\cite{Dolag:2004kp}.  The spectrum for ``MHD(x)'' (``Model~3x'') model, with the mean field of 100~fG (0.1 fG), is close to that of the uniform 100~fG (0.1~fG) field.  One can also understand this from general arguments~\cite{Essey:2009zg,Essey:2009ju,Essey:2010er}.  Thus, we modeled the IGMF as a characteristic strength organized into cubes of a characteristic correlation length each with a random direction.
Of course, if the distribution of matter along the line of sight is known, e.g., from Lyman-$ \alpha $ forest in the direction to some blazars, one can further improve our inferences of IGMFs.  

From the available sources, we chose a subset based on three conditions: (a) the source must be observed at energies above 1~TeV, (b) the distance to the source must be relatively large, and (c) the source must show no temporal variability above 1~TeV.  The latter condition is an necessary but not sufficient condition that secondary gamma rays dominate the observed signal, since primary gamma rays from blazars are predicted to show temporal variability.  Based on these criteria, we chose three sources: 1ES 0229+200 ($z=0.14$), 1ES1101-232 ($z=0.186$), and 1ES0347-121 ($z=0.188$), each of which has been observed by HESS~\cite{Aharonian:2007wc,Aharonian:2007nq,Aharonian:2007tc}. Temporal variability has been measured above the ~200~GeV energy threshold of the HESS array for 1ES1101 and
   1ES0347 \cite{Zech:ICRC}. 
\begin{figure}
  \begin{center}
\begin{tabular}{c}
 \includegraphics[height=0.25 \textheight]{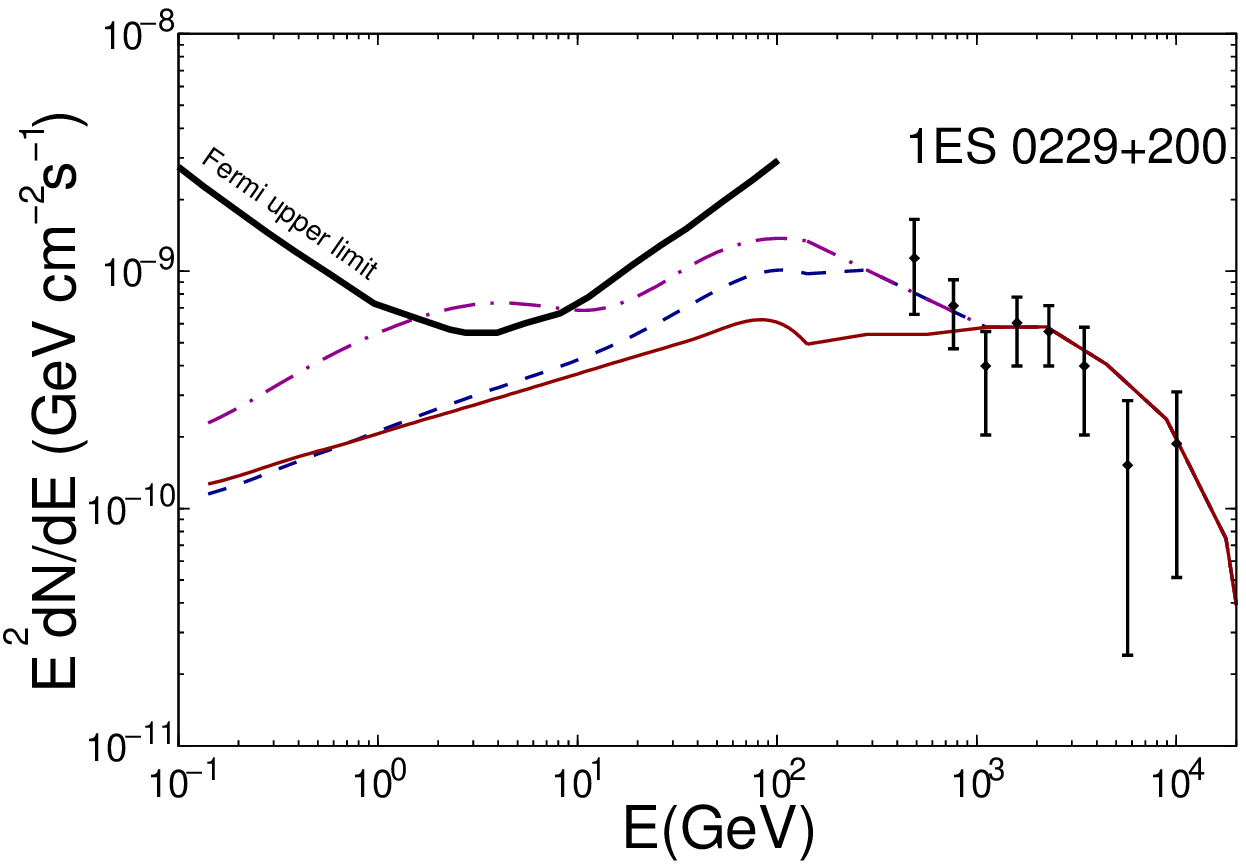}\\
 \includegraphics[height=0.25 \textheight]{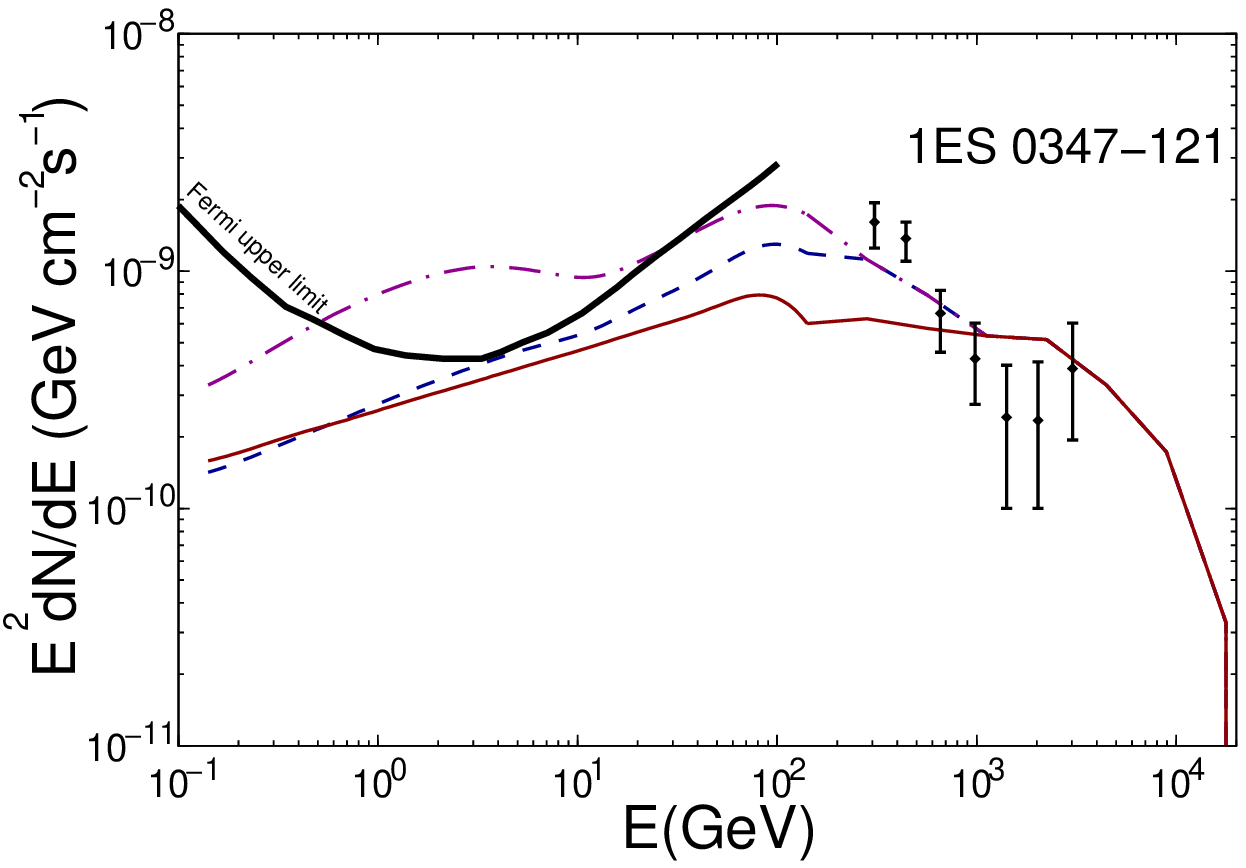}\\  
 \includegraphics[height=0.25 \textheight]{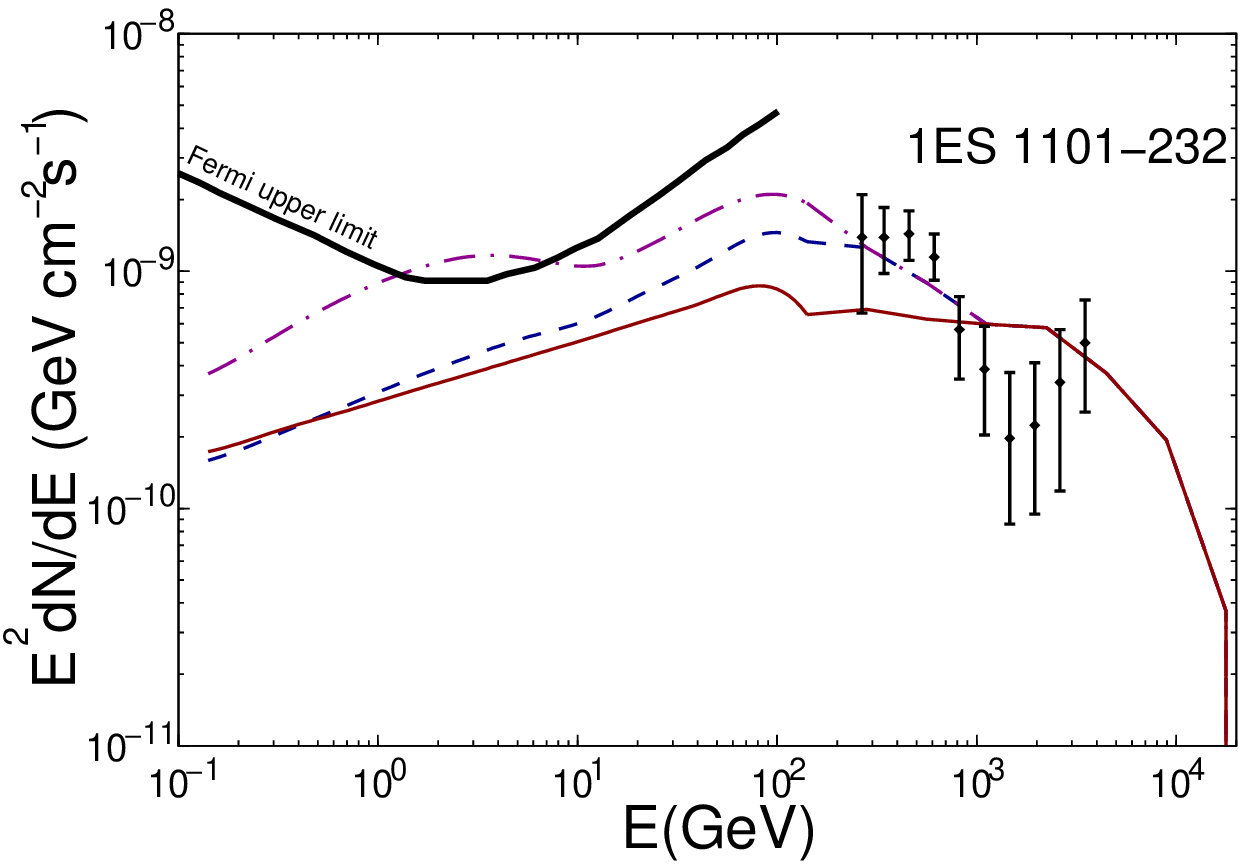}
\end{tabular}
\end{center}
\caption{ 
Photon spectra calculated numerically and normalized to fit HESS data points for 1ES0229+200 ($z=0.14$)~\cite{Aharonian:2007wc}, 1ES1101-232 ($z=0.186$)~\cite{Aharonian:2007nq}, and 1ES0347-121 ($z=0.188$)~\cite{Aharonian:2007tc}.  The Fermi upper limits at lower energy (thick solid line) were derived from the data in Ref.~\cite{Neronov:1900zz}.  The three model predictions are for primary photons and $B=10^{-18}$~G (upper, dash-dotted line), $B=10^{-15}$~G (middle, dashed line), and B~$=10^{-13}$~G (lower, thin solid line). The PSF of Fermi is folded into the signal below 100~GeV and the PSF of HESS is folded into the signal above 100~GeV.  Here we assumed gamma-ray spectrum at the source $\propto E^{-1.75} $ with a 100~TeV high-energy cutoff. 
}
  \label{fig:spechi}
\end{figure}

We consider a range of viable EBL models.  A lower limit on the EBL density is set by galaxy counts~\cite{Xu:2000ss}.   The upper limit is less known.  Although gamma-ray data have been used to exclude higher EBL models such as Stecker et al.~\cite{Stecker:2005qs}, it was recently pointed out that the inclusion of secondary gamma rays from VHE gamma rays and cosmic rays significantly improves the fits to the data for these ``high EBL'' models~\cite{Essey:2010er}.  Therefore, we include the model of Stecker et al.~\cite{Stecker:2005qs} as the upper limit (for UV EBL).  A number of other models fall between these two limits~\cite{Salamon:1997ac}. EBL models with a higher level of UV background radiation are difficult to reconcile with a limit based on gamma-ray burst data~\cite{LAT:2010kz}.

We shall first discuss the case of primary gamma rays only. Following most recent papers on IGMFs, we will assume that the blazar spectra 
follow a power law $dN/dE \sim \rm E^{-\Gamma}$ with $\Gamma \geq 1.5$~\cite{1981MNRAS.196..135P}.  However  harder spectra are also possible~\cite{Stecker:2007zj}. 
Furthermore, it is possible that the intrinsic spectra do not conform to a single power law.  Deviations from single power law may affect the limits on IGMFs derived under the assumption of a 
power-law spectrum.  We  consider the range $\Gamma = 0.5-2$, which encompasses a broad variety of models. We also allow the high-energy cutoff of the power law spectrum to vary in a wide range, $10-10^4$~TeV.

We will consider magnetic fields ranging from $10^{-18}$~G to $10^{-11}$~G. The IGMF correlation lengths are also not known; we assume it is in the range 0.01--1~Mpc and use 1~Mpc as a reference point in Fig.~1.

We fit the calculated spectrum to the data obtained by HESS Atmospheric Cherenkov Telescope (ACT) at high energy, perform statistical analysis to calculate 95\% confidence intervals, and we demand that the low-energy prediction not overshoot the Fermi upper limit.  The spectra obtained for 1ES~0229+200 ($z=0.14$), 1ES~1101-232 ($z=0.186$), and 1ES~0347-121 ($z=0.188$),  are shown in Fig.~1 for IGMFs in the range $10^{-18}-10^{-13}$~G, with a correlation length $\lambda_c = 1$~Mpc. The results for a source with intrinsic spectrum $\Gamma=1.75$, a 100~TeV high-energy cutoff and the ``low'' EBL model~\cite{Xu:2000ss}, shown in Fig.~1, illustrate the effects of IGMF on the spectrum.  For larger IGMFs, more of the secondary gamma rays arrive outside the cone determined by the angular resolution of the instrument, with lower energy gamma rays arriving at larger angles.   This causes an energy dependent drop in the spectrum.

As one can see from comparing  $10^{-15}$~G and $10^{-13}$~G curves in Fig.~1, the wash-out of the signal due to magnetic deflections causes a break in the spectrum at some energy, which is a function of the magnetic field.  The following semi-quantitative analytical estimate may be helpful.  The magnetic deflections depend on the energy~\cite{Neronov:2009gh}: $$\Theta_{\rm defl}\simeq 0.1^\circ (1+z)^{-2}(\tau_\theta / 5)^{-1}(E_\gamma / 0.1~TeV)^{-1}(B / 10^{-15}~G),$$ for $\rm \lambda_c\gg D_e$, where $\tau_\theta$ is the optical depth of the primary gamma ray and $\rm D_e$ the electron cooling distance. For the energies considered here, $\tau_{\theta} \propto E_\gamma^{-1/2}$, and for inverse-Compton (IC) cooling $E_\gamma \propto E_{\gamma 0}^2$, where $E_{\gamma 0}$ is the energy of the primary photon, and $n_\gamma \propto E_\gamma^{-1/2}$, where $n_{\gamma}$ is the number of photons produced from IC cooling. Thus we expect $\Theta_{\rm defl} \propto B/E_{\gamma 0}^{3}$. Let us consider the flux of gamma rays that contribute to the point image (determined by a fixed instrumental resolution):  $\Theta_{\rm defl} < \Theta_{PSF}$ implies $  E_{\gamma 0} > {\rm const} \times B^{1/3}$. Thus for an intrinsic power-law spectrum with index $\Gamma$ we expect the point flux   
$$ F 
\propto \frac{1}{E_{\gamma}^{1/2}}\int^\infty_{B^{1/3}}\frac{dN}{dE_{\gamma 0}}\sim E_{\gamma}^{-1/2}\int^\infty_{B^{1/3}}E_{\gamma 0}^{-\Gamma}\sim \frac{1}{E_{\gamma}^{1/2}B^{(\Gamma-1)/3}}.  
$$
This relation implies the approximate scaling $B^{2(\Gamma-1)} E^3 = {\rm const}$ for the position of the break in the spectrum.   By comparing the $10^{-15}$~G and $10^{-13}$~G curves in Fig.~\ref{fig:spechi} we see that this approximation works reasonably well.  Of course, this discussion is admittedly simplistic: the relative contributions of primary and secondary gamma rays, and some other effects were neglected.  In deriving the actual limits we use the results of detailed numerical Monte Carlo for propagation, shower development, and the effects of magnetic fields. 

For angular resolution of about $0.1^\circ$ we expect magnetic fields
larger than about 1~fG to wash out the signal observed by the ACTs,
while fields lower than a fG would boost the signal in the Fermi energy
range. This allows us to use the combination of ACT and Fermi data to
set both upper and lower limits on the IGMF. A lower limit can be found
by requiring that the IGMF is large enough to wash out enough of the GeV
signal so that the observed spectrum is below the Fermi 95\% CL upper
limits for the source. An upper limit can be found by requiring that the IGMF be small enough
to avoid washing out the TeV signal.
More specifically, we combined the ACT data for these three blazars and used 25 data points shown in Fig.~1.
Then for each model with given $B$ and $\Gamma$, we have three
independent parameters, i.e., normalizations (or luminosities) of the
spectra of three blazars. The number of degrees of freedom of the fit is 22.
If the value of $\chi_{\rm min}^2$ for each model specified with fixed
$B$ and $\Gamma$ is larger than 33.9, then such a model is acceptable at the level of 
only 5\% probability. This way, we set excluded regions (or upper limits on $B$) at 95\% CL.

Our results for pure gamma-ray sources are summarized in Fig.~2. The results depend on the choice of EBL model and high energy cutoff. We show both the ``low'' EBL model, based on lower limits from galaxy counts, and the ``high'' EBL model~\cite{Stecker:2005qs}, as well as high-energy cutoff values of 20~TeV and 100~TeV.

Let us now include the the cosmic ray component~\cite{Essey:2009zg,Essey:2009ju,Essey:2010er}.
Now the observed gamma-ray spectrum does not depend on the intrinsic spectrum of gamma rays, and it is also practically independent of the spectral 
properties of cosmic rays, such as the high energy cutoff and the cosmic ray spectral index (under some very mild assumptions)~\cite{Essey:2009zg,Essey:2009ju,Essey:2010er}. 
Therefore, instead of the exclusion regions shown in Fig.~2, we find the following bounds (for $\lambda_c=1$~Mpc):
\begin{eqnarray}
3\times 10^{-16}~{\rm G} & <  B < &\rm 3\times 10^{-14}~{\rm G \ (High \ EBL) }\nonumber\\
 1\times 10^{-17}~{\rm G} & < B < &\rm 8\times 10^{-16}~{\rm G \ (Low \ EBL)} \label{IGMF_limits}
\end{eqnarray}
Of course it is possible, and likely, that the observed spectrum is a combination of secondary particles from gamma rays and cosmic rays emitted by the source. In this case the most conservative allowed range for IGMF would be the union of the gamma-ray and cosmic-ray allowed ranges.


\begin{figure*}
\begin{tabular}{cc}
\includegraphics[width=0.45\textwidth]{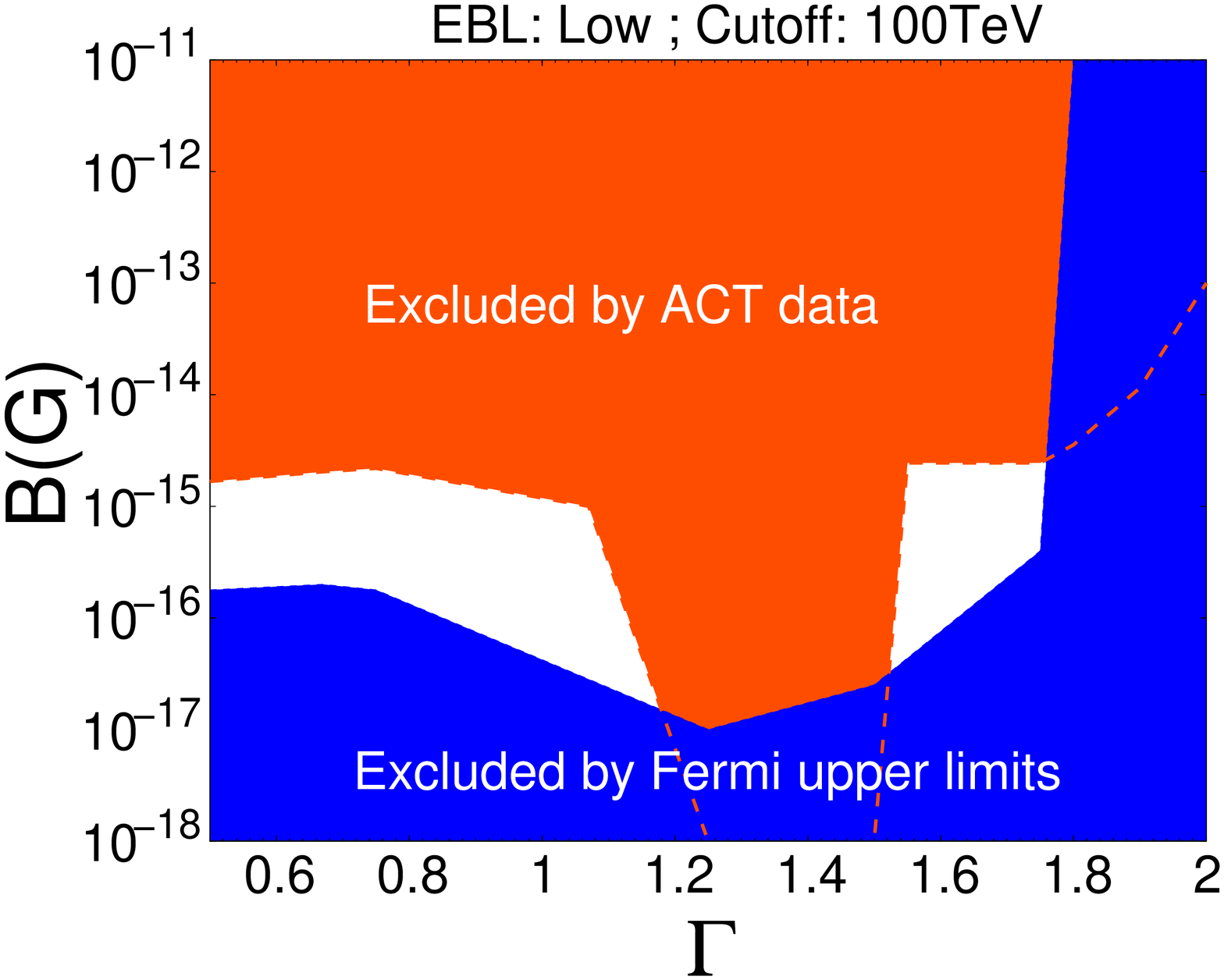}
& 
\includegraphics[width=0.45\textwidth]{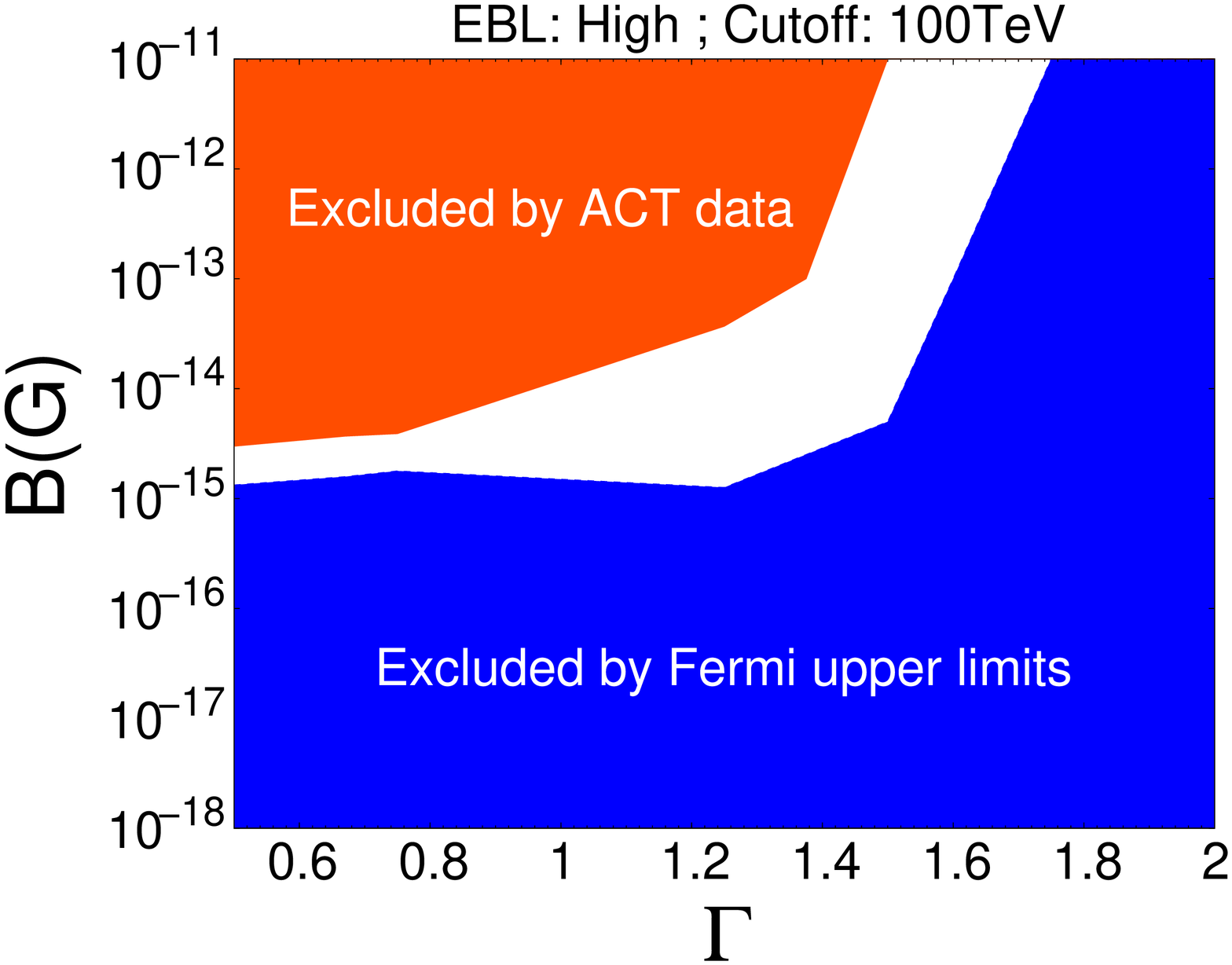}
\\ & \\
\includegraphics[width=0.45\textwidth]{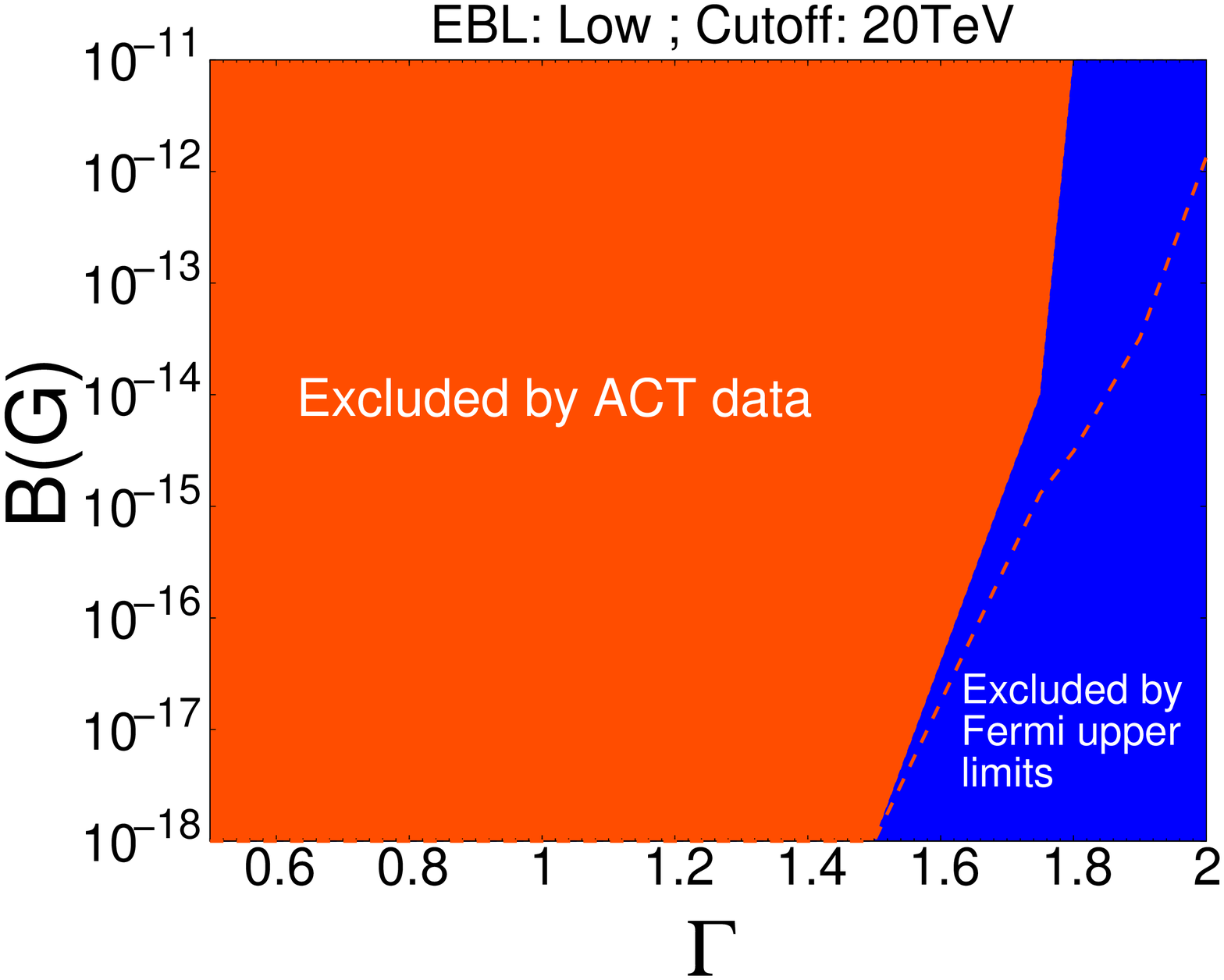}
& 
\includegraphics[width=0.45\textwidth]{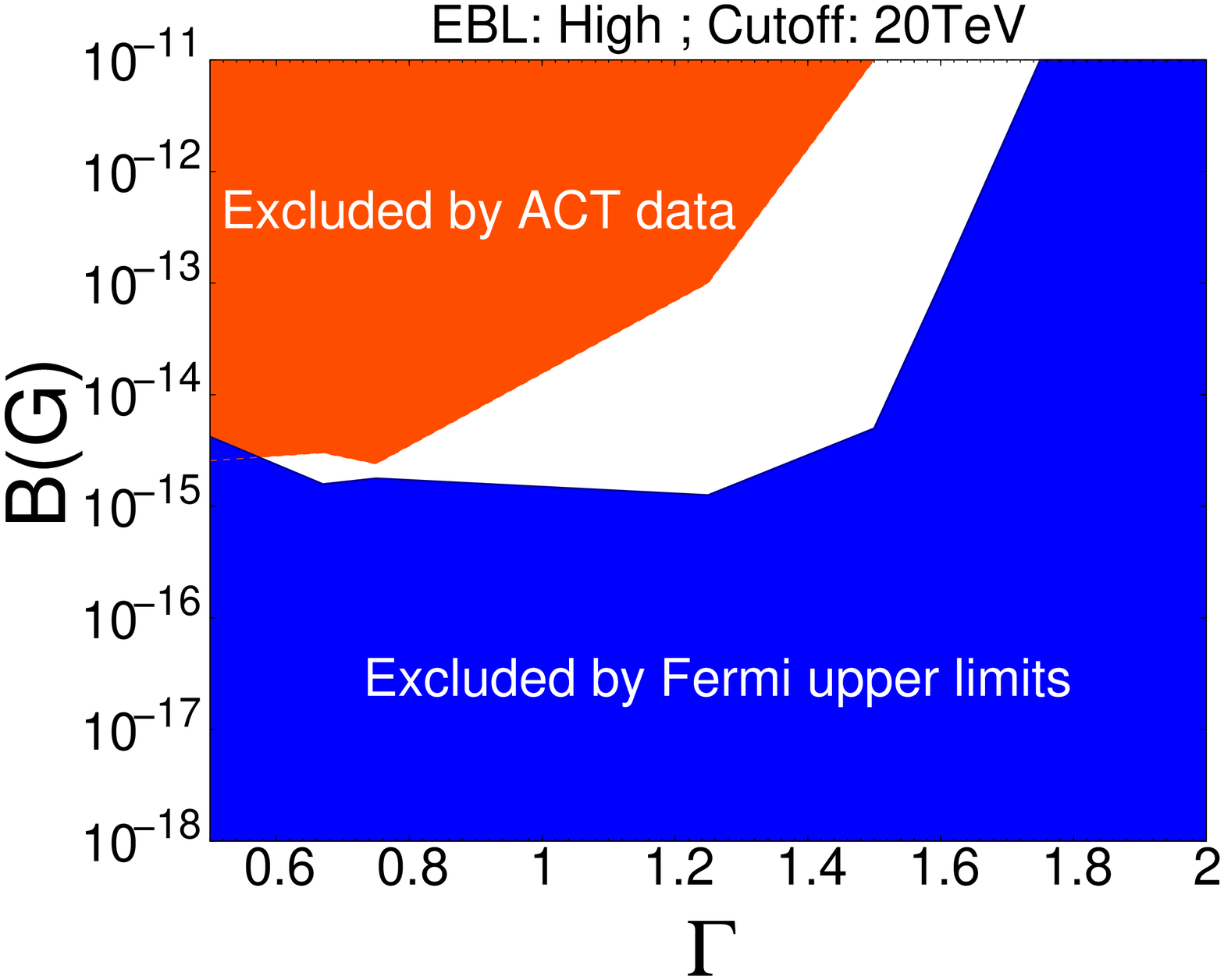}
\end{tabular}
\caption{Limits on the magnitude of the IGMF and intrinsic spectral
 index based on the combined fit for spectra of 1ES 0229+200, 1ES
 1101-232, and 1ES 0347-121, assuming that only gamma rays are produced
 at the source (no cosmic ray contribution).   Shaded regions are
 excluded at 95\% CL from either Fermi or ACT data. The limits are shown
 for ``low'' and ``high'' EBL models and for two values of the
 high-energy cutoff: 100~TeV and 20~TeV.  IGMF correlation length is
 assumed to be 1~Mpc.
}
\label{fig:photon_spectra}
\end{figure*}

The limits in Fig.~2 imply $\Gamma \le 1.8$ for the set of blazars we chose, which is in good agreement with mean spectral index of 1.9 reported by Fermi for nearby sources. 

The limits presented in Fig.~2 are subject to some caveats. First, we have assumed that our three sources had similar intrinsic spectra, which need not be the case.  Second, as discussed above, the likely contribution of protons along the line of sight~\cite{Essey:2009zg,Essey:2009ju,Essey:2010er} can reduce the limits to the model-independent form shown in Eq.~\ref{IGMF_limits}.  Finally, time variability of the primary gamma ray sources can affect our conclusions~\cite{Dermer:2010mm}.  

Following the insightful discussion of Dermer et al.~\cite{Dermer:2010mm}, it can be shown that the typical time delays for secondary photons are
\begin{equation}
\Delta t \simeq 2\times 10^6 \lambda_{100} \left(\frac{B_{-15}}{E_\gamma/10~\rm{GeV}}\right)^2~\rm{yrs}, 
\label{eq:timedelays}
\end{equation}
where $\lambda_{100}\sim1$ for the Stecker et al EBL model, $B_{-15}$ is the IGMF in units of $10^{-15}$~G. If the source is active for a timescale that is shorter than these delays, 
the observed signal is suppressed by a factor $(t_{\rm active}/\Delta t)$ where $t_{\rm active}$ is the time that the source was active for. This suppression mimics the effects of 
a stronger magnetic field, and, therefore, the predicted limit on the IGMF would be suppressed by the a factor
\begin{equation}
\kappa \simeq \left(\frac{t_{\rm active}}{2\times 10^6~\rm{yrs}}\right)^{1/2}\left(\frac{10^{-15}~\rm{G}}{B_{\rm limit}}\right)\left(\frac{E_\gamma}{10~\rm{GeV}}\right), 
\label{eq:timedamp}
\end{equation}
where $B_{\rm limit}$ is either the upper or lower limit shown in Fig.~2. 

The observations of 1ES 0229+100 by HESS~\cite{Aharonian:2007wc} and VERITAS~\cite{Perkins:VERITAS} were separated by roughly 3 years and show no evidence of significant variability. 
Variability may be present during the times when the source was not observed, or at a level below the instrumental sensitivity. We assume $t_{\rm active}\sim 3$~yrs as a lower limit on the time the source was active. The most restrictive energy for the lower limits is near $E_\gamma \sim 10$~GeV and most of the lower limits fall in the range $10^{-15}-10^{-17}$~G which means the most conservative lower limits would be suppressed by a factor of $10-10^3$. For upper limits, the most restrictive energy is at $E_\gamma\sim 1$~TeV and since most upper limits fell into the range $10^{-13}-10^{-15}$~G we expect the upper limits to also be lowered by roughly $10-10^3$. Thus, the overall effect of including a short activity time is to lower the 95\% confidence intervals shown in Fig.~2 and our robust $10^{-17}$~G lower limit would be lowered to $10^{-18}$~G in the most conservative case.

It is important to distinguish between the time scale of the source activity and short-term variability.  A variability of the source would not lead to a suppression discussed above because signals of variable activity would undergo significant delays in the IGMF and would be observed as a constant source with the time averaged luminosity of the flares. Thus, if the source is active on a timescale of $10^6$~yrs, the limits in Fig.~2 are valid, even in the case of variability on shorter time scales. 

\begin{figure}
\begin{center}
\includegraphics[width=\textwidth]{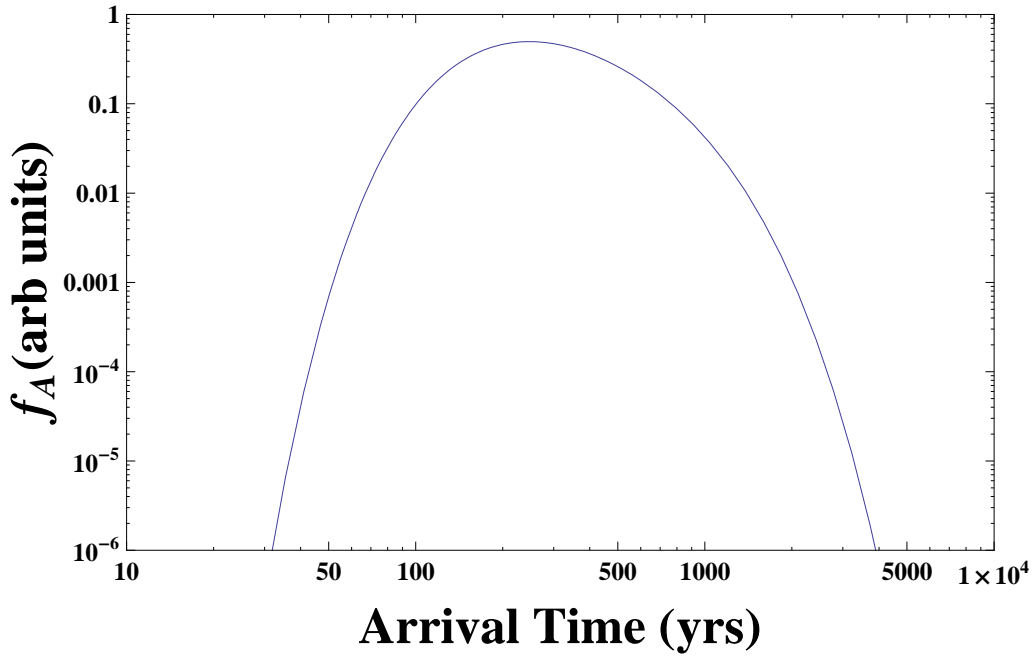}
\hspace{0.5cm}
\caption{
Arrival time probability distribution in arbitrary units for primary cosmic rays including only delays at the source. A cosmic ray source of $10^{10}$~GeV protons was used in a source 100~kpc wide with $B=10^{-6}$~G and $l_c=0.1$~kpc.}
\label{prosource1e6}
\end{center}
\end{figure}

The effect the source activity--dormancy time scale is much weaker in the case of the cosmic ray limits, because the delays in the stronger magnetic fields near the source can 
wash out variability on even longer time scales. The magnetic fields within the host galaxies containing source are likely to be of the order of $1~\mu G$ which can lead to significant delays. Fig.~\ref{prosource1e6} shows the delays for a 100~kpc wide source of $10^{10}$~GeV protons with $B=1~\mu \rm{G}$ and $l_c=0.1$~kpc. It can be seen that the delays are significantly longer than the observed variability of blazars and thus we expect any intrinsic variability to be washed out. Such a source appears as a constant, long-lived source, and any time delays incurred in the host galaxy are not  measurable.   Of course, there is a significant uncertainty in the magnetic field distribution around the source, in particular, in the direction of the jet, and the values assumed here are taken 
solely to illustrate the idea.

The effect of the source variability would be to damp the observed power of the source by a factor of 
\begin{equation}
 f_{damp}\sim N_{\rm active}\frac{t_{\rm active}}{t_{\rm delay}}, 
\end{equation}
where $t_{\rm delay}$ is the typical proton delay at the source, $t_{\rm active}$ is the typical time the source is active or flaring and $N_{\rm active}$ is the number of times the source is active in the 
time period $t_{\rm delay}$. This damping will not be a significant effect, especially since the typical deflections at the source will not be enough to affect the beaming factors.
 
Using Eq.~(\ref{eq:timedamp}) and these typical time delays we see that the model independent lower limit of $10^{-17}$~G would remain unchanged and the upper limit of $3\times 10^{-14}$~G would be lowered by roughly one order of magnitude. Thus the limits we provided for cosmic rays are still the most conservative limits even including all timing uncertainties. Thus, for the case of cosmic rays dominating the signal, we can report a model-independent 95\% CL interval for the IGMF of
\begin{equation}
1\times 10^{-17}~\rm G < B < 3\times 10^{-14}~G.
\end{equation}

Our results represent the first 95\% confidence interval measurements of the IGMF based on the spectral data alone.  The method used here to probe the IGMF will become more powerful in the future. Future observations of distant blazars will not only improve the limits on the IGMF, but also constrain AGN and EBL models.  The measured values of IGMFs can be used to distinguish between the cosmological and astrophysical scenarios for the origin of all astrophysical magnetic fields. In addition, this information will lead to a better understanding of gamma rays and cosmic rays, as well as the properties of AGN and of universal photon backgrounds. 

The authors thank F.~Aharonian, J.~Beacom, P.~Blasi, C.~Dermer, S.~Razzaque, and E.~Waxman for helpful comments. 
The work of W.E. and A.K. was supported by DOE grant DE-FG03-91ER40662 and NASA ATP grant NNX08AL48G. 
The work of S.A. was supported by Japan Society for Promotion of Science.

\end{document}